\documentclass[prl, showpacs, twocolumn, floatfix]{revtex4}

\usepackage{graphicx}
\usepackage{amsmath, amsfonts, amssymb, bm}
\usepackage[]{psfrag}

\psfragscanoff
\begin{document}

\title{Particle physics with a laser-driven positronium atom}
\author{Carsten \surname{M\"{u}ller}}
\email{c.mueller@mpi-hd.mpg.de}
\author{Karen Z. \surname{Hatsagortsyan}}
\email{k.hatsagortsyan@mpi-hd.mpg.de}
\author{Christoph H. \surname{Keitel}}
\email{keitel@mpi-hd.mpg.de}
\affiliation{Max-Planck-Institut f\"ur Kernphysik, Saupfercheckweg 1, D-69117 Heidelberg, Germany}

\date{\today}

\begin{abstract}
A detailed quantum-electrodynamic calculation of muon pair creation in 
laser-driven electron-positron collisions is presented. The colliding particles 
stem from a positronium atom exposed to a superintense laser wave of linear 
polarization, which allows for high luminosity. The threshold laser intensity 
of this high-energy reaction amounts to a few 10$^{22}$~W/cm$^2$ in the 
near-infrared frequency range. The muons produced form an ultrarelativistic, 
strongly collimated beam, which is explicable in terms of a classical simple-man's 
model. Our results indicate that the process can be 
observed at high positronium densities with the help of present-day laser technology.
\end{abstract}
 
\pacs{13.66.De, 
41.75.Jv, 
36.10.Dr  
}

\maketitle
Laser fields can pave new ways to high-energy physics \cite{GeV,report,Kirk,help}.
By laser-matter interactions in the MeV regime, fundamental nuclear physics 
processes were realized in experiment \cite{nuclear} and new laser-induced 
nuclear phenomena have been predicted theoretically \cite{Thomas}. While inside 
the most powerful laser fields presently available \cite{GeV}, electrons 
temporarily acquire ponderomotive energies in the GeV range. Relativistic 
rectification techniques of the laser field based on laser-plasma interaction 
enable real particle acceleration and extraction of monoenergetic electron 
(ion) beams up to GeV (MeV) energies \cite{accel}. 
However, the electron's temporary energy gain in the laser beam can directly be 
exploited without any rectification technique when particle collisions followed 
by particle reactions happen {\it inside} the laser beam \cite{Kirk,collider}.

A further advantage of laser fields can be utilized when the interaction
with single atoms rather than a plasma is considered. Then, apart from the high 
energies achievable, lasers can be used to generate well-controlled coherent 
collisions of the atomic constituent particles at microscopic impact parameters,
which can lead to enormous luminosities \cite{Corkum,collider}. 
An atomic species of particular interest 
in this context is positronium (Ps), the bound state of an electron and a positron, 
since it exhibits unique dynamic properties in a strong laser field \cite{Ps}: 
After instantaneous ionization the leptons oscillate in opposite directions along
the laser electric field, which leads to periodic $e^+e^-$ collisions since both
particles experience an identical ponderomotive drift motion due to their equal charge-to-mass ratios. Note that in ordinary atoms, the magnetically induced
drift in the laser propagation direction suppresses recollisions at high intensities
\cite{report}.

In the following we consider laser-induced muon pair creation from electron-positron 
annihilation (see Fig.\,1). 
In contrast to the usual setup 
of colliding particle beams, however, we assume that an initially quiescent Ps 
atom is exposed to a superintense laser wave of linear polarization and the muons 
are produced in the laser-driven $e^+e^-$ collisions described above. 
The corresponding reaction rate is evaluated within the framework of laser-dressed
quantum electrodynamics by employing relativistic Volkov states \cite{LL}.
We show that the minimal laser intensity to ignite the process is determined by 
the relation $eA\ge Mc^2$, with electron charge $-e$, muon rest energy 
$Mc^2\approx 106$~MeV, and amplitude of the laser's vector potential $A$,
as the collision energy comes from the transversal motion.
For a laser wavelength $\lambda =1$ $\mu$m, the threshold intensity amounts to $5.5\times 10^{22}$ W/cm$^2$ which is almost reached by present most powerful 
laser systems \cite{Uggerhoj}. Such a superintense laser wave supplies the electron-positron system with an energy of $2Mc^2$ in their center-of-mass (c.m.) frame,
which coincides with the energetic threshold for muon production in the field-free case.
Note that the required intensity thus lies several orders of magnitude below the critical
value of $2.3\times 10^{29}$ W/cm$^2$ where $e^+e^-$ pair creation from vacuum and
other vacuum nonlinearities in laser fields are expected to appear \cite{GeV,report}.
The rates for the laser-induced decay ${\rm Ps}\to\mu^+\mu^-$ are shown to render experimental observation feasible at high Ps density ($\sim 10^{18}$ cm$^{-3}$ 
\cite{BEC}) and laser repetition rate ($\sim {\rm Hz}$ \cite{GeV}).

\begin{figure}[b]
\begin{center}
\resizebox{4.5cm}{!}{\includegraphics{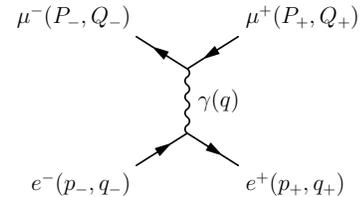}}
\caption{\label{diagramm} Feynman graph for muon pair creation from electron-positron annihilation in a laser field. The arrows are labeled by the particle's free momenta ($p_\pm, P_\pm$) outside and the effective momenta ($q_\pm, Q_\pm$) inside the laser field.} 
\end{center} 
\end{figure}

Lepton-lepton interactions in a laser field have been studied in detail
by the examples of laser-assisted $e^-e^-$ (M{\o}ller) and $e^+e^-$ 
(Bhabha) scattering \cite{Moller}. In these cases the laser wave served as 
a background field which modified the field-free properties of the scattering 
reaction. Contrary to that, in the present scenario the laser 
wave is playing a vital role in initiating a process that would not take place in 
the absence of the laser field.

Within a strong-field approximation, the amplitude for $e^+e^-\to\mu^+\mu^-$
from a laser-driven Ps atom can be written as a superposition integral \cite{muon}
\begin{eqnarray}
\label{SPs}
\mathcal{S}_{\rm Ps} = {1\over\sqrt{\mathcal{V}}}\int{d^3p\over(2\pi)^3}\,
           \tilde\Phi(\mbox{\boldmath$p$})\,\mathcal{S}_{e^+e^-}
\end{eqnarray}
with a normalization volume $\mathcal{V}$ and the Fourier transform $\tilde\Phi(\mbox{\boldmath$p$})$ of the Ps ground-state wave-function which 
depends on the relative momentum $\mbox{\boldmath$p$}$ of the $e^+e^-$ two-body system. $\tilde\Phi(\mbox{\boldmath$p$})$ represents the probability amplitude 
for finding a positron of momentum $\mbox{\boldmath$p$}_+ = \mbox{\boldmath$p$}$ 
and, correspondingly, an electron of momentum 
$\mbox{\boldmath$p$}_- = -\mbox{\boldmath$p$}$ within the Ps ground-state 
wave-packet. Furthermore,
\begin{eqnarray}
\label{See}
\mathcal{S}_{e^+e^-} &=& -{\rm i}\alpha\int d^4x \, d^4y\,
\overline\Psi_{p_+,s_+}(x)\gamma^\mu\Psi_{p_-,s_-}(x)\nonumber\\ 
&\times& \!\!\! D_{\mu\nu}(x-y)
\overline\Psi_{P_-,S_-}(y)\gamma^\nu\Psi_{P_+,S_+}(y)
\end{eqnarray}
denotes the amplitude for the process $e^+e^-\to\mu^+\mu^-$ in a laser 
wave (cf. Fig.\,1). Note that we use relativistic units with $\hbar=c=1$.
In Eq.\,(\ref{See}), $\alpha$ is the finestructure constant, $\Psi_{p_\pm,s_\pm}$ 
($\Psi_{P_\pm,S_\pm}$) are the Volkov states \cite{LL} of the $e^\pm$ ($\mu^\pm$)
depending on the free four-momenta $p_\pm$ ($P_\pm$) and spin states $s_\pm$ ($S_\pm$) 
of the particles, and $D_{\mu\nu}$ is the coordinate-space representation 
of the free photon propagator. The integrations in Eq.\,(\ref{See}) are performed 
in the usual way by Fourier series expansion where the Fourier coefficients are 
given by generalized Bessel functions \cite{genbes}. At the electron-positron 
vertex they read $J_n(u,v)$, with the index $n$ corresponding to the number of 
laser photons absorbed or emitted, and the arguments given by
\begin{equation}
\label{arg}
u = {eA(\epsilon p_-)\over (kp_-)}-{eA(\epsilon p_+)\over (kp_+)}\,,\
v = -{e^2A^2\over 8}\left[{1\over (kp_-)} + {1\over (kp_+)}\right].
\end{equation}
Here, $k^\mu=\omega(1,0,0,1)$, ${\mathcal A}^\mu(x)=\epsilon^\mu A\,{\rm cos}(kx)$ with $\epsilon^\mu=(0,1,0,0)$, and $\xi = eA/(m\sqrt{2})$ are the laser's wave four-vector, 
four-potential, and normalized vector potential, respectively. Note that 
$v\approx v_0\equiv -m\xi^2/(2\omega)$ is practically constant for the relevant 
values of $|\mbox{\boldmath$p$}|\sim m\alpha$. Similar generalized Bessel functions 
$J_N(U,V)$ appear at the muon-antimuon vertex, where $U,V$ are given by 
Eq.\,(\ref{arg}) with the corresponding momenta of muon and antimuon.
Due to the properties of the Bessel functions, the typical photon number is 
$n\approx -m\xi^2/\omega$ so that the intermediate photon four-momentum satisfies
\begin{eqnarray}
\label{q}
q^2 = (q_+ + q_- - nk)^2 \approx 4m_\ast^2-4n\omega m \approx 8m^2\xi^2,
\end{eqnarray}
implying that the process proceeds nonresonantly (i.e., $q^2\neq 0$) \cite{Moller}.
Here, $q_\pm$ denotes the effective $e^\pm$ momenta inside the laser field and
$m_\ast = m(1+\xi^2)^{1/2}$ is the laser-dressed mass \cite{LL}. By virtue of an 
addition theorem for Bessel functions \cite{genbes}, Eq.\,(\ref{See}) adopts the 
structure
\begin{eqnarray}
\label{See2}
\mathcal{S}_{e^+e^-} &=& -{{\rm i}\alpha\over 8m^2\xi^2}
\sum_{r\ge r_{\rm min}} J_r(U-u,V-v)\,T_r\nonumber\\
& &\times\, \delta(q_+ + q_- - Q_+ - Q_- + rk)
\end{eqnarray}
where $r\equiv N-n$ is the total number of absorbed laser photons, 
$r_{\rm min}\equiv(M^2-m^2)/(\omega m)$, and $T_r$ are rather complicated 
functions of the particle momenta and laser parameters. The main $p$ dependence 
in Eq.\,(\ref{SPs}) is contained in oscillatorily damped integrals of the form 
\begin{eqnarray}
\label{I}
I_r &\equiv& \int {d^3p\over(2\pi)^3} \tilde\Phi(\mbox{\boldmath$p$})J_r(U-u,V-v)
\nonumber\\
&\approx& {J_r(U,\Delta V)\over\sqrt{(2\pi)^3a_0^3}}
\left({\omega\over m\alpha^2\xi}\right)^{2/3}
{2^6(\xi\Delta V)^{1/3}\over\alpha^{5/3}|v_0|}\,, 
\end{eqnarray}
with $\Delta V\equiv V-v_0$ and the Ps Bohr radius $a_0$.
The remaining $p$ dependence of $\mathcal{S}_{e^+e^-}$, neglected in Eq.\,(\ref{I}), 
is very weak because of the nonrelativistic momenta contained in the Ps ground state 
\cite{muon}. According to the factor $[\omega/(m\alpha^2\xi)]^{2/3}\sim[a_0/(\alpha\xi\lambda)]^{2/3}\ll 1$ 
in Eq.\,(\ref{I}), 
the average over the Ps momentum distribution leads to destructive interference 
of the partial waves constituting the bound state, which can equivalently be described 
as wave packet spreading [cf. Eq.\,(\ref{RPs3}) below]. 

The total rate for laser-driven Ps decay into muons is found from the square of 
the amplitude (\ref{SPs}), averaged and summed over the particle spins and 
integrated over the outgoing muon momenta:
\begin{eqnarray}
\label{RPs}
R_{\rm Ps} &\approx&
  {\alpha^2\over 2^8 m^4\xi^4}
  \int {d^3P_+\over (2\pi)^3} \int {d^3P_-\over (2\pi)^3}
  \sum_{s_\pm,S_\pm} \sum_{r\ge r_{\rm min}} |I_r T_r|^2\nonumber\\
 & &  \times\, \delta(q_+ + q_- - Q_+ - Q_- + rk).
\end{eqnarray} 

Via the energy-momentum conserving $\delta$-function in Eq.\,(\ref{RPs}),
the muon kinematics can be analyzed. One finds that the laser intensity parameter 
needs to satisfy
\begin{eqnarray}
\label{ximin}
\xi\ge\xi_{\rm min}\equiv {M\over m\sqrt{2}}\approx 150.
\end{eqnarray} 
It is remarkable that Eq.\,(\ref{ximin}) agrees with the naive estimate of the 
threshold intensity given above, although the muons have to be created with their laser-dressed mass $M_\ast=M[1+(m\xi/M)^2]^{1/2}$ which is significantly larger 
than their bare mass $M$. The energetic difference $\Delta M^2\equiv 2(M_\ast^2-M^2)$
is supplied by the laser photons absorbed at the scattering event.
Close to threshold, the muons have typical momenta of 
\begin{eqnarray}
\label{typicalP}
P_x \approx M\,,\ \ P_y \approx 0\,,\ \ P_z \approx M^2/m. 
\end{eqnarray}
The total number of absorbed laser photons is $r\approx 2M^2/(\omega m)$, which is 
of order 10$^{10}$ at $\omega = 1$ eV and in agreement with the energy
conservation law: $r\omega \approx 2(P_0-m)$. Note that, even at threshold,
the muons are highly relativistic so that their life time in the lab frame 
is increased from $2.2~\mu$s to $\sim {\rm ms}$ due to time dilation.

The typical muon momenta in Eq.\,(\ref{typicalP}) can also be deduced from a 
classical simple-man's model. To this end, assume that at the reaction
threshold ($\xi=\xi_{\rm min}$), the muons are created with zero momentum in 
the (primed) $e^+e^-$ c.m. frame: $\mbox{\boldmath$P$}'(\tau_0)=0$,
where $\tau$ denotes the laser phase. The solution of the classical 
equations of motion with this initial condition reads
$P'_x(\tau) = e[A(\tau)-A(\tau_0)]$, $P'_y(\tau) = 0$, 
$P'_z(\tau) = {e^2\over 2M}[A(\tau)-A(\tau_0)]^2$. 
After the laser field has passed, it becomes $P'_x = M$, $P'_y = 0$, 
$P'_z = M/2$. Note here that the vector potential at the moment of creation 
attains its maximum value $A(\tau_0)=A$, since the laser's electric 
field is zero at the recollision time. The Lorentz transformation to the lab frame 
then yields the momentum components of Eq.\,(\ref{typicalP}).

From Eq.\,(\ref{RPs}) one can derive, similarly to Ref.\,\cite{muon}, 
an approximate formula for the total rate of muon production from a single 
laser-driven Ps atom
\begin{eqnarray}
\label{RPs2}
R_{\rm Ps} \approx {2^7\over\pi^2}{\alpha^2\over m^2\xi^2 a_0^3}
  \sqrt{1-{\xi_{\rm min}^2\over\xi^2}}
  {1\over\xi}\left({a_0\over\alpha\xi\lambda}\right)^{\! 3}
  \left({m\over\omega\xi^4}\right)^{\! 1/3}.
\end{eqnarray}
The analytical estimate (\ref{RPs2}) is confirmed by direct numerical evaluation 
of Eq.\,(\ref{RPs}), as shown in Fig.\,2. The high-order Bessel function in 
Eq.\,(\ref{I}) was evaluated with the help of a suitable asymptotic expansion.
Equation~(\ref{RPs2}) has an intuitive interpretation in terms of the cross 
section for field-free $e^+e^-\to\mu^+\mu^-$ \cite{Peskin}
\begin{eqnarray}
\label{sigma}
\sigma = {4\pi\over 3}{\alpha^2\over E_{\rm cm}^2}\sqrt{1-{4M^2\over E_{\rm cm}^2}}
\left(1+{2M^2\over E_{\rm cm}^2}\right)
\end{eqnarray}
and the $e^\pm$ 
wave packet size which is spreading in the laser field due to quantum mechanical
dispersion: 

In our case, the c.m. energy in Eq.\,(\ref{sigma}) is $E_{\rm cm}\approx 2\sqrt{2}m\xi$. The wave-packet spreading can be estimated as follows: in the $e^+e^-$ c.m. frame, the initial momentum spread is $\delta p^{\prime}\sim 1/a_0$. The spreading during
the recollision time $t_{r}^{\prime}$ can be estimated as 
$\delta x^{\prime}\sim \delta y^{\prime} \sim \delta z^{\prime} \sim 
\delta p^{\prime}t_{r}^{\prime}/m.$
Due to relativistic time dilation, the oscillation period in the c.m. frame
is largely enhanced so that the recollision time at $\xi \gg 1$ equals $t_{r}^{\prime}\sim 4\pi \gamma_z/\omega$, with the Lorentz factor 
$\gamma_z\sim \xi$. At the collision event, the wave packet size has thus spread to 
$\delta x^{\prime}\delta y^{\prime}\delta z^{\prime}\sim (\alpha\xi\lambda)^3$.
The corresponding particle current density leads to a reaction rate
$R_{\rm Ps}^\prime \sim \sigma/(\alpha\xi\lambda)^3$. By boosting this rate
to the lab frame, we obtain
\begin{eqnarray}
\label{RPs3}
R_{\rm Ps} \sim {\sigma\over\xi(\alpha\xi\lambda)^3}
\end{eqnarray}
which agrees with Eq.\,(\ref{RPs2}) up to a factor of order unity.
Thus, our rigorous quantum-electrodynamic calculation demonstrates for the 
first time that simple rate estimates for (nonresonant) nuclear or particle 
reactions in strong laser fields \cite{Kirk,Corkum,collider}, based on 
semi-classical arguments and field-free cross sections, are indeed reliable.

The information given above on the muon kinematics and the typical photon 
numbers is also corroborated by numerical calculations. Figure~3 shows partial 
creation rates, i.e. the rates $R_r$ for muon production by absorption of 
a certain number $r$ of laser photons, their sum yielding the total 
reaction rate: $R_{\rm Ps}=\sum_r R_r$. The partial rates reflect the energy 
distribution of the created particles and agree with the analytical estimates.

\begin{figure}[t]
\begin{center}
\includegraphics*[width=0.87\columnwidth]{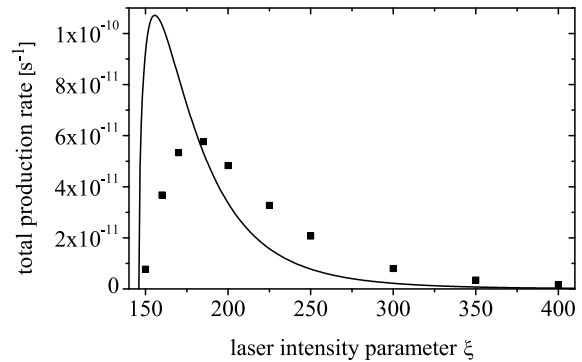}
\caption{\label{rate} Total rate for laser-driven $\mu^+\mu^-$ creation from
a Ps atom as a function of the intensity parameter ($\omega=1$ eV). 
The solid line shows the analytical estimate in Eq.\,(\ref{RPs2}); 
the black squares result from numerical calculations based on Eq.\,(\ref{RPs}).} 
\end{center} 
\end{figure}

\begin{figure}[b]
\begin{center}
\includegraphics*[width=0.87\columnwidth]{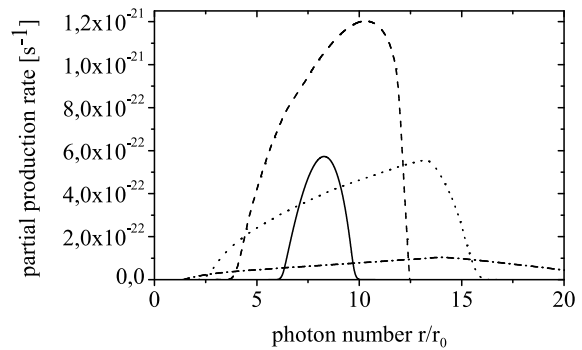}
\caption{\label{rate} Partial rates for the laser-driven reaction 
${\rm Ps}\to\mu^+\mu^-$ as a function of the absorbed photon number $r$ 
for intensity parameters of $\xi=150$ (solid line), 170 (dashed line), 
200 (dotted line), and 250 (dash-dotted line). The photon number is 
given in units of $r_0\equiv m\xi^2/(2\omega)$, where $\omega=1$ eV.} 
\end{center} 
\end{figure}

The muon production rate from a single Ps atom attains a maximum value of 
order 10$^{-10}$ s$^{-1}$ at a laser intensity of $10^{23}$ W/cm$^2$ (see Fig.\,2). 
Since high laser intensities 
are obtained by short, tightly focussed pulses, the experimental observation
of the predicted process requires dense Ps targets. The highest Ps density
achieved so far is of order $10^{15}$ cm$^{-3}$ \cite{BEC}; recent
proposals aim at $10^{18}$ cm$^{-3}$ with regard to antihydrogen experiments
\cite{antihydrogen} or generation of a Ps Bose-Einstein condensate \cite{BEC}. 
At the latter density, a typical laser focal volume $V_{\rm f}\approx (10\lambda)^3$ would
contain about $10^9$ Ps atoms, so that the total rate roughly amounts to
1 s$^{-1}$ which still seems too small to be measured in view of the 
short duration of strong laser pulses ($\sim$ fs\,$-$\,ns).
However, the wave packet spreading which has a detrimental impact on the reaction 
rate, can be controlled by applying a second counterpropagating laser beam 
\cite{collider,Corkum}. This kind of laser configuration is realized, e.g. by the
Astra Gemini system at the Rutherford Appleton Laboratory, U.K. or the
JETI "photon collider" at the University of Jena, Germany \cite{RAL}. 
Then the spreading factor $[a_0/(\alpha\xi\lambda)]^3$ in Eq.\,(\ref{RPs2}) is 
absent and the muon yield could be increased by $\sim 10$ orders of magnitude.
Under these circumstances, with kJ laser systems of high repetition rates
($\sim {\rm Hz}$) \cite{GeV} the observation of laser-driven muon creation
from Ps is expected to be feasible.

The problem of wave packet spreading can be avoided by laser-driven
collisions of {\it free} electrons and positrons, e.g. in an $e^\pm$ plasma.
Then integrals like the one in Eq.\,(\ref{I}) are absent.
Based on Eq.\,(\ref{See}), the muon creation rate in this situation is 
approximately obtained as follows
\begin{eqnarray}
\label{Rfree}
R_{e^+e^-} \approx {1\over 2^3\pi^2}{\alpha^2\over m^2\xi^4}
\sqrt{1-{\xi_{\rm min}^2\over\xi^2}}\ {N_+N_-\over V_{\rm int}}.
\end{eqnarray}
Here, $N_\pm$ is the number of $e^\pm$ contained in the interaction volume
$V_{\rm int}$. By employing realistic values of $V_{\rm int}=(10\lambda)^3$ and
$N_\pm=10^7$ (corresponding to a plasma density of $10^{16}$ cm$^{-3}$ 
\cite{antihydrogen}), we obtain $R_{e^+e^-}\lesssim 10^{-2}$ s$^{-1}$, which is 
much smaller than $R_{\rm Ps}$ in the crossed-beams setup. This corroborates 
the high luminosity achievable from Ps by laser-guided microscopic collisions.

Finally we note that the process ${\rm Ps}\to\mu^+\mu^-$ is heavily suppressed in a circularly polarized laser field \cite{muon}. This is 
expressed by a damping factor $[a_0/(\lambda\xi)]^4\sim 10^{-25}$ arising from 
the classical motion of the $e^+$ and $e^-$ that co-rotate in the polarization plane 
at a macroscopic distance $\sim\lambda\xi$. Contrary to that, in linearly polarized
fields the classical $e^\pm$ trajectories periodically meet and the rate damping
results from quantum mechanical dispersion. Equation~(\ref{RPs3}) thus describes muon creation in the collision of two extended but localized quantum wave packets. The reaction rate, being several orders of magnitude larger than for circular laser polarization, clearly indicates the distinct physical nature of the collision process.


In conclusion, we have studied the high-energy reaction of muon pair creation by a 
nonrelativistic Ps atom subject to a superintense laser field of linear polarization. 
Due to the threshold intensity $\approx 10^{23}$ W/cm$^2$, the experimental investigation 
of this process which involves billions of photons would naturally be a goal for the next 
generation of high-power laser systems \cite{ELI}. The measurement, though challenging,
is facilitated by the ultrarelativistic kinematics of the created muons and requires 
high Ps densities which might be provided by other fields of physics. The results of the present study also hold for similar high-energy reactions like, e.g., pion or rho-meson production. Microscopic $e^+e^-$ collisions arising from Ps in strong laser fields thus have the potential of paving an alternative route to laser particle physics.


\begin{thebibliography}{99}

\bibitem{GeV} G. A. Mourou, T. Tajima, and S. V. Bulanov, Rev. Mod. Phys. {\bf 78}, 
309 (2006); M. Marklund and P. K. Shukla, {\it ibid.} {\bf 78}, 591 (2006).

\bibitem{report} Y. I. Salamin {\it et al.}, Phys. Rep. {\bf 427}, 41 (2006).

\bibitem{Kirk} 
K. T. McDonald and K. Shmakov, Phys. Rev. STAB {\bf 2}, 121301 (1999).

\bibitem{help} 
T. Tajima and G. Mourou, {\it ibid.} {\bf 5}, 031301 (2002);
K. Nakajima, AIP Conf. Proc. {\bf 737}, 614 (2004).

\bibitem{nuclear} H. Schwoerer, J. Magill, and B. Beleites (Eds.), {\it Lasers 
and Nuclei} (Springer, Heidelberg, 2006);
T. Ditmire {\it et al.}, Nature (London) {\bf 398}, 489 (1999);
D. Umstadter, J. Phys. D {\bf 36}, R151 (2003);
K. W. D. Ledingham, P. McKenna, and R. P. Singhal, Science {\bf 300}, 1107 (2003).

\bibitem{Thomas} T. J. B\"urvenich, J. Evers, and C. H. Keitel, Phys. Rev. Lett. 
{\bf 96}, 142501 (2006); 
V. Yu. Bychenkov {\it et al.}, Pis'ma Zh. Eksp. Teor. Fiz. {\bf 74}, 664 (2001) 
[JETP Lett. {\bf 74}, 586 (2001)]. 

\bibitem{accel} 
S. P. D. Mangles {\it et al.}, Nature {\bf 431}, 535 (2004); 
C. G. R. Geddes {\it et al.}, {\it ibid.} {\bf 431}, 538 (2004); 
J. Faure {\it et al.}, {\it ibid.} {\bf 431}, 541 (2004);
W. P. Leemans {\it et al.}, Nature Phys. {\bf 2}, 696 (2006);
B. M. Hegelich {\it et al.}, Nature {\bf 439}, 441 (2006);
H. Schwoerer {\it et al.}, {\it ibid.} {\bf 439}, 445 (2006);
T. Toncian {\it et al.}, Science {\bf 312}, 410 (2006).

\bibitem{collider} K. Z. Hatsagortsyan, C. M\"uller, and C. H. Keitel, 
Europhys. Lett. {\bf 76}, 29 (2006).

\bibitem{Corkum} N. Milosevic, P. B. Corkum, and T. Brabec, Phys. Rev. Lett. 
{\bf 92}, 013002 (2004); 
see also G. R Mocken and C. H. Keitel, J. Phys. B {\bf 37}, L275 (2004).

\bibitem{Ps} B. Henrich, K. Z. Hatsagortsyan, and C. H. Keitel, Phys. Rev. Lett. 
{\bf 93}, 013601 (2004).

\bibitem{LL} V. B. Berestetskii, E. M. Lifshitz, and L. P. Pitaevskii, 
{\it Relativistic Quantum Theory} (Pergamon, Oxford, 1971).

\bibitem{Uggerhoj} Already with existing laser systems, Ps atoms can be exposed 
to such high intensities if, instead of a fixed target, a relativistic 
Ps beam is used [see U. I. Uggerh{\o}j, Phys. Rev. A {\bf 73}, 052705 (2006)].

\bibitem{BEC} 
D. B. Cassidy {\it et al.}, Phys. Rev. Lett. {\bf 95}, 195006 (2005).

\bibitem{Moller} 
O. I. Denisenko and S. P. Roshchupkin, Laser Phys. {\bf 9}, 1108 (1999);
P. Panek, J. Z. Kaminski, and F. Ehlotzky, Phys. Rev. A {\bf 69}, 013404 (2004),
and references therein.

\bibitem{muon} C. M\"uller, K. Z. Hatsagortsyan, and C. H. Keitel, 
Phys. Rev. D {\bf 74}, 074017 (2006).

\bibitem{genbes} H. R. Reiss, Phys. Rev. A {\bf 22}, 1786 (1980).

\bibitem{Peskin} M. E. Peskin and D. V. Schroeder, {\it An Introduction 
to Quantum Field Theory} (Addison-Wesley, Reading, 1995).

\bibitem{antihydrogen} P. Perez and A. Rosowsky, Nucl. Instrum. Meth. Phys. Res. A
{\bf 545}, 20 (2005).

\bibitem{RAL} For current information see http://www.clf.rl.ac.uk
and http://www.physik.uni-jena.de/\raisebox{-0.9ex}{\~{ }}ioq, respectively.


\bibitem{ELI} See, e.g., the proposal on the Extreme Light Infrastructure 
(ELI) available at http://www.eli-laser.eu.

\end{thebibliography}
\end{document}